\IfFileExists{IEEEtran.cls}
  {\documentclass[conference]{IEEEtran}}
  {\documentclass[10pt,twocolumn]{article}\usepackage[margin=0.75in]{geometry}}
\usepackage[utf8]{inputenc}
\usepackage[T1]{fontenc}
\usepackage{lmodern}
\usepackage{amsmath,amssymb}
\usepackage{graphicx}
\usepackage{booktabs}
\usepackage{xcolor}
\usepackage{tikz}
\usetikzlibrary{positioning,arrows.meta,fit,backgrounds}
\usepackage{url}
\usepackage[hidelinks]{hyperref}
\newcommand{\code}[1]{\texttt{#1}}

\title{Graphs Have Structure: Declarative Manifolds, 3D Z-Order Monoid Roll-ups, and Zero-Simulation Visual Analytics in a Graph DBMS}
\author{%
  \IEEEauthorblockN{Madhulatha Mandarapu}
  \IEEEauthorblockA{Samyama.ai\\madhulatha@samyama.ai}
  \and
  \IEEEauthorblockN{Sandeep Kunkunuru}
  \IEEEauthorblockA{Samyama.ai\\sandeep@samyama.ai}
}

\begin{document}
\maketitle

\begin{abstract}
Every real-world graph is embedded in an ambient structural manifold: proteins fold in Euclidean space ($\mathbb{R}^3$), telecom networks sit on a planetary surface ($\mathbb{S}^2 \times \mathbb{R}$), and ontologies and calendars form discrete subsumption posets ($(V, \sqsubseteq)$). Yet property graph databases treat node coordinates as opaque payload scalars, leave spatial proximity edges unindexed, and force visualizers into $O(V^2)$ client-side force-directed simulations that destroy true geometry and truncate hierarchy roll-ups.

We elevate ambient manifold structure to a first-class primitive in Samyama Graph. First, we introduce a declarative and auto-detected structural convention contract (\code{samyama.structure/1}) persisted in \code{.sgsnap} headers that binds node labels to Euclidean 3D, Geo-Spherical, Causal Spacetime, or Poset Hierarchy manifolds. Second, we show that both discrete posets (via Order-Embedded Hierarchy $[\text{tin}, \text{tout}]$ DFS intervals) and continuous 3D manifolds (via a 63-bit 3D Morton Z-order Structural-Geometric Index, SGI) reduce regional aggregation to 1D rank-interval monoid folds answered directly from index-resident Fenwick trees (\code{SUM}, \code{COUNT}) and Segment Trees (\code{MIN}, \code{MAX}). At $100{,}000$ 3D nodes, SGI builds in $28.14\text{ ms}$ and answers Morton-interval \code{SUM} and \code{MAX} roll-ups in $0.048\,\mu\text{s}$ and $0.104\,\mu\text{s}$---$8{,}345.1\times$ and $3{,}824.4\times$ faster than property scans. Third, we introduce 3D voxel-hashed virtual contact edges weighted by Topological--Geometric Stress $\sigma(u,v) = d_{\text{graph}}(u,v) / d_{\text{manifold}}(u,v)$ to isolate long-range tertiary folds, plus manifold-guided $A^*$ shortest-path search that cuts explored states by up to $3.80\times$ over Dijkstra. Fourth, we replace client layout simulation with server-computed 3D Cone-Tree/Ribbon coordinates and a non-WebGL 2.5D Isometric projection with painter's-order depth sorting and sub-millisecond Level-of-Detail (LoD) macro-cuts.
\end{abstract}


\section{Introduction}\label{sec:intro}

Property graph databases model data as a combinatorial pair $G = (V, E)$ adorned with key-value property maps. In this abstraction, a graph has adjacency but no intrinsic geometry: the coordinates $(x, y, z)$ of an amino acid residue in a Protein Data Bank (PDB) structure~\cite{berman2000}, the $(\text{lat}, \text{lon}, \text{alt})$ of a cellular base station, and the taxonomic depth of a concept in the NCBI Taxonomy~\cite{ncbitaxonomy} or GeoNames~\cite{geonames} are stored as uninterpreted floating-point or integer payload properties.

Treating ambient structure as opaque payload creates four systemic failures across the query engine and its visual analytics stack:

\begin{enumerate}
\item \textbf{Regional roll-ups degenerate into linear scans.} Asking for the sum, minimum, or maximum of a measure (such as crystallographic B-factor, cell traffic volume, or descendant taxon count) over a 3D spatial region or taxonomic subtree forces the query engine to scan candidate nodes and dereference property storage, rather than answering the aggregate from an index.
\item \textbf{Implicit spatial adjacency is either unindexed or quadratic.} In structural biology and wireless infrastructure, the most consequential interactions are \emph{non-bonded contacts}---pairs of nodes $(u, v)$ that are far apart along the covalent or routing backbone ($d_G(u,v) \gg 1$) but lie within a physical cutoff radius $\epsilon$ in ambient space ($d_{\mathcal{M}}(u,v) \le \epsilon$). Materializing all pairs within $\epsilon$ across multiple cutoffs bloats edge storage, while computing them on the fly without a 3D index costs $O(V^2)$.
\item \textbf{Graph algorithms run blind to the manifold.} Pathfinding (\code{shortestPath}) and community detection execute uninformed combinatorial searches (BFS or Dijkstra) that expand wavefronts radially in hop space, ignoring admissible geometric lower bounds available from the node coordinates.
\item \textbf{Visualizers fabricate $O(V^2)$ hairballs and truncate hierarchy roll-ups.} When a browser client such as Samyama Insight fetches a subgraph of $250$ to $2{,}000$ nodes, it runs an iterative force-directed layout~\cite{fruchterman1991} that destroys true 3D molecular or geospatial geometry. Worse, when rendering a hierarchy from a client-loaded subgraph, client-side aggregation can only sum over the loaded subset, misreporting the size and roll-up of every collapsed ancestor near the root.
\end{enumerate}

\paragraph{The thesis.} Every real-world graph is embedded in an ambient structural manifold $\mathcal{M}$---discrete or continuous---and declaring that manifold to the database unifies indexing, regional roll-up, structure-guided algorithms, and zero-simulation visual analytics under a single abstraction.

Building on our prior work on Order-Embedded Hierarchy indexing (OEH) for subsumption posets~\cite{mandarapu2026oeh}, we observe a fundamental algebraic duality between discrete hierarchies and continuous 3D space. In a tree or near-tree poset, a depth-first numbering maps every subtree to a contiguous 1D rank interval $[\text{tin}(u), \text{tout}(u)]$. In continuous 3D space, quantizing $(x, y, z)$ into a 63-bit 3D Morton Z-order code~\cite{morton1966} maps octree cells and Morton ranges into contiguous 1D rank intervals $[\text{rank}(m_{\text{lo}}), \text{rank}(m_{\text{hi}})]$. Consequently, \emph{both discrete posets and continuous 3D manifolds can be aggregated by the exact same index-resident monoid engines}: a Fenwick tree~\cite{fenwick1994} for invertible group operations (\code{SUM}, \code{COUNT}) and a Segment Tree for idempotent semilattice operations (\code{MIN}, \code{MAX}).

\paragraph{Contributions.} We design, implement, and evaluate this architecture inside the Rust-based Samyama Graph engine~\cite{mandarapu2026samyama}:

\begin{itemize}
\item \textbf{C1 --- Declarative \& Auto-Detected Structural Conventions (\S\ref{sec:model}):} We introduce the \code{samyama.structure/1} specification persisted in \code{.sgsnap} snapshot headers, together with a schema-free property-signature detector that binds node labels to four manifold classes: \code{Euclidean3d}, \code{GeoSpherical}, \code{HierarchyPoset}, and \code{CausalSpacetime}.
\item \textbf{C2 --- Unified 1D Rank-Interval Monoid Indexing via OEH and SGI (\S\ref{sec:indexing}):} We pair OEH (extended with a 4-way structural probe including \code{NearTree} exception lists and Segment Tree \code{MIN}/\code{MAX} roll-ups) with the \emph{Structural-Geometric Index} (\textbf{SGI}), a 63-bit 3D Morton Z-order index with index-resident Fenwick and Segment Tree roll-ups.
\item \textbf{C3 --- Structure-Guided Graph Operators (\S\ref{sec:algo}):} We implement 3D voxel-hashed virtual contact extraction, \emph{Topological--Geometric Stress} $\sigma(u,v) = d_G(u,v) / d_{\mathcal{M}}(u,v)$ to isolate long-range tertiary folds and routing detours, and manifold-guided $A^*$ shortest-path search~\cite{hart1968}.
\item \textbf{C4 --- Zero-Simulation 3D \& Non-WebGL 2.5D Isometric Visual Cues (\S\ref{sec:visual}):} We expose engine-level HTTP endpoints (\code{/api/hierarchies} and \code{/api/structure/cues}) that return authoritative whole-graph roll-ups, 3D Cone-Tree~\cite{robertson1991} and polymer ribbon coordinates, and a deterministic non-WebGL 2.5D Isometric projection with painter's-algorithm depth sorting and sub-millisecond Level-of-Detail (LoD) macro-cuts.
\end{itemize}

\begin{figure*}[t]
\centering
\begin{tikzpicture}[
  box/.style={draw, rounded corners=3pt, minimum height=0.8cm, align=center, font=\footnotesize, fill=gray!8},
  hdr/.style={draw, rounded corners=3pt, minimum height=0.8cm, align=center, font=\footnotesize\bfseries, fill=blue!8},
  eng/.style={draw, rounded corners=3pt, minimum height=0.9cm, align=center, font=\footnotesize, fill=green!8},
  api/.style={draw, rounded corners=3pt, minimum height=0.85cm, align=center, font=\footnotesize, fill=orange!10},
  arr/.style={-{Stealth[length=2.2mm]}, thick}
]
  \node[hdr, text width=3.4cm] (snap) at (0, 0) {\code{.sgsnap} Snapshot Header \\ \& Auto-Probe (\code{detect\_conventions})};

  \node[box, text width=3.5cm] (oeh) at (4.6, 1.15) {\textbf{Discrete Poset (\code{OEH})} \\ \code{Tree} / \code{NearTree} / \code{Chain} \\ DFS Interval $[\text{tin}(u), \text{tout}(u)]$};
  \node[box, text width=3.5cm] (sgi) at (4.6, -1.15) {\textbf{Continuous 3D (\code{SGI})} \\ \code{Euclidean3d} / \code{GeoSpherical} \\ 63-bit Morton $[\text{rank}(m_{\text{lo}}), \text{rank}(m_{\text{hi}})]$};

  \node[eng, text width=3.8cm] (monoid) at (9.3, 1.15) {\textbf{1D Rank-Interval Monoids} \\ \code{Fenwick}: \code{SUM}, \code{COUNT} ($O(\log n)$) \\ \code{SegmentTree}: \code{MIN}, \code{MAX} ($O(\log n)$)};
  \node[eng, text width=3.8cm] (algo) at (9.3, -1.15) {\textbf{Structure-Guided Operators} \\ 3D Voxel Contacts + Stress $\sigma(u,v)$ \\ Manifold-Guided $A^*$ Pathfinding};

  \node[api, text width=3.6cm] (cues) at (14.0, 0) {\textbf{Zero-Simulation Visual Cues} \\ 3D Cone-Tree / Ribbon Coords \\ Non-WebGL 2.5D Isometric + LoD};

  \draw[arr] (snap.east) -- (oeh.west);
  \draw[arr] (snap.east) -- (sgi.west);
  \draw[arr] (oeh.east) -- (monoid.west);
  \draw[arr] (sgi.east) -- (monoid.west);
  \draw[arr] (sgi.east) -- (algo.west);
  \draw[arr] (monoid.east) -- (cues.west);
  \draw[arr] (algo.east) -- (cues.west);
\end{tikzpicture}
\caption{Architecture of Manifold-Embedded Property Graphs in Samyama Graph. Both discrete subsumption posets (OEH) and continuous 3D manifolds (SGI) linearize structural regions into 1D rank intervals backed by the same Fenwick and Segment Tree monoid engines, powering structure-guided operators and zero-simulation 3D and 2.5D visual analytics.}
\label{fig:arch}
\end{figure*}
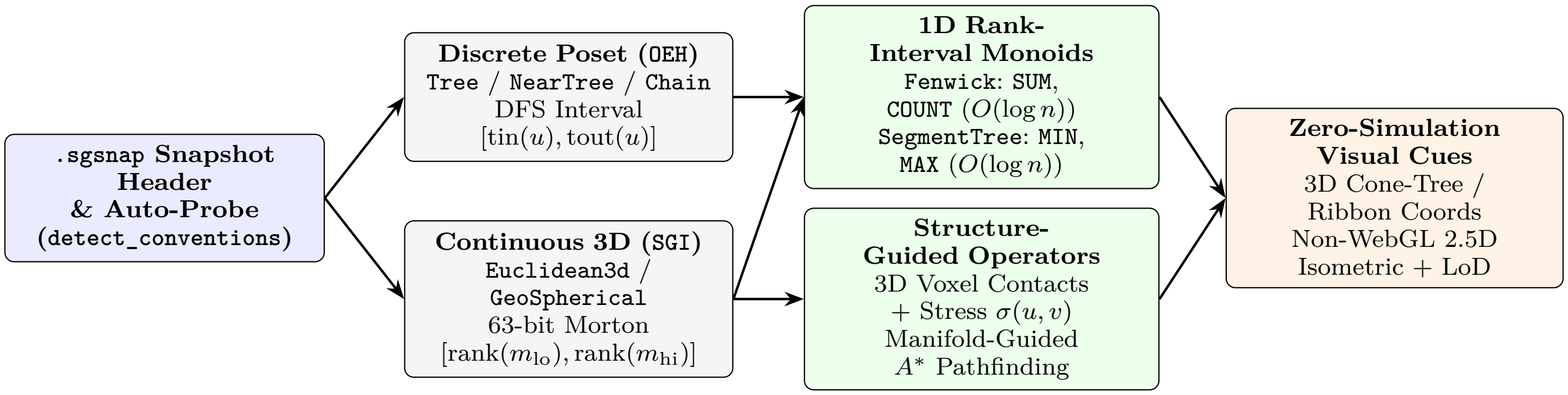

\section{The Manifold-Embedded Property Graph Model}\label{sec:model}

\subsection{Formal Definition}

A \emph{Manifold-Embedded Property Graph} is a tuple
\[
\mathcal{G} = (V, E, \lambda, \mathcal{M}, \pi_{\mathcal{M}}, \mu)
\]
where $(V, E, \lambda)$ is a labeled property graph, $(\mathcal{M}, d_{\mathcal{M}})$ is an ambient metric or partially ordered manifold, $\pi_{\mathcal{M}} : V_{\mathcal{M}} \subseteq V \to \mathcal{M}$ is a partial embedding mapping nodes of a target label to points in $\mathcal{M}$, and $\mu : V_{\mathcal{M}} \to M$ maps nodes to elements of a commutative monoid $(M, \oplus, \mathbf{0})$.

Unlike a purely combinatorial graph where distance is solely shortest-path hop count $d_G(u,v)$, every embedded pair $u, v \in V_{\mathcal{M}}$ possesses two distances:
\begin{enumerate}
\item \textbf{Topological distance} $d_G(u,v)$, the shortest path length along explicit edges $E$; and
\item \textbf{Manifold distance} $d_{\mathcal{M}}(\pi_{\mathcal{M}}(u), \pi_{\mathcal{M}}(v))$, evaluated directly in the ambient space $\mathcal{M}$.
\end{enumerate}
The interplay between $d_G$ and $d_{\mathcal{M}}$ governs both structural indexing and structure-guided graph algorithms.

\subsection{Canonical Manifold Classes}

Figure~\ref{fig:arch} illustrates how Samyama Graph binds node labels and edge types to four canonical manifold classes (\code{ManifoldSpec} in \code{src/index/structure/mod.rs}):

\paragraph{1. \code{Euclidean3d} ($\mathcal{M} = \mathbb{R}^3$).}
Nodes carry three floating-point coordinate properties $(x, y, z)$ with a physical unit (e.g., \code{angstrom} for biomolecular structures or \code{m} for LiDAR/CAD digital twins), an optional monotonic sequence property $s(u)$ (such as \code{residue\_seq} along a polypeptide chain), a chain grouping property (such as \code{chain\_id}), and a default spatial contact cutoff $\epsilon > 0$. The metric is Euclidean $L_2$:
\[
d_{\mathbb{R}^3}(u, v) = \|\pi_{\mathcal{M}}(u) - \pi_{\mathcal{M}}(v)\|_2.
\]

\paragraph{2. \code{GeoSpherical} ($\mathcal{M} = \mathbb{S}^2 \times \mathbb{R}$).}
Nodes carry geodetic latitude and longitude $(\phi, \lambda)$ in degrees (default CRS \code{EPSG:4326}) and an optional altitude $h$ in meters. The ambient distance combines great-circle Haversine distance $d_{\text{hav}}((\phi_u, \lambda_u), (\phi_v, \lambda_v))$ on Earth radius $R_\oplus = 6{,}371{,}000\text{ m}$ with vertical separation $\Delta h$:
\[
d_{\mathbb{S}^2 \times \mathbb{R}}(u, v) = \sqrt{d_{\text{hav}}(u,v)^2 + (h_u - h_v)^2}.
\]
For 3D rendering and voxel hashing, $(\phi, \lambda, h)$ is projected to Earth-Centered, Earth-Fixed (ECEF) Cartesian coordinates $(r\cos\phi\cos\lambda,\, r\cos\phi\sin\lambda,\, r\sin\phi)$ with $r = R_0 + \alpha h$.

\paragraph{3. \code{HierarchyPoset} ($\mathcal{M} = (V, \sqsubseteq)$).}
Generated by a transitive covering relation $R_{\sqsubseteq} \subseteq E$ (e.g., \code{IS\_A}, \code{PART\_OF}, \code{SUBCLASS\_OF}) maintained by an \code{OehIndex}. Here the structural predicates are subsumption ($u \sqsubseteq v$), Lowest Common Ancestor ($\text{LCA}(u,v)$), and poset depth $\text{depth}(u)$.

\paragraph{4. \code{CausalSpacetime} ($\mathcal{M} = \mathbb{R}^2 \times \mathbb{T}$).}
Nodes carry spatial coordinates $(x, y)$ and a timestamp $t$, supporting causal cone queries where $t_u \le t_v$ and $\|(x_u - x_v, y_u - y_v)\|_2 \le c_{\max}(t_v - t_u)$.

\subsection{Declarative Persistence in \code{.sgsnap} \& Auto-Detection}

To ensure that datasets ship with their structural semantics intact, we extend the \code{.sgsnap} binary snapshot header (\code{SnapshotHeader}) with an additive \code{structures: Vec<StructuralConvention>} field. When empty, the field is omitted via \code{skip\_serializing\_if = "Vec::is\_empty"}, preserving byte-for-byte backward compatibility with existing snapshots.

When a graph is loaded without explicit conventions, \code{detect\_conventions(\&GraphStore)} performs a bounded sample probe ($k \le 32$ nodes per label) checking: (i) all active \code{OehIndex} instances $\to$ \code{HierarchyPoset}; (ii) coordinate triples $(\code{x},\code{y},\code{z})$ or $(\code{ca\_x},\code{ca\_y},\code{ca\_z})$ plus sequence properties $(\code{residue\_seq}, \code{seq\_id})$ $\to$ \code{Euclidean3d} with preset \code{protein\_ribbon\_3d} (default cutoff $\epsilon = 8.0\,\text{\AA}$); and (iii) bounded latitude/longitude pairs $(\phi \in [-90^\circ, 90^\circ], \lambda \in [-180^\circ, 180^\circ])$ $\to$ \code{GeoSpherical}.

\section{Unified 1D Rank-Interval Monoid Indexing}\label{sec:indexing}

\subsection{Rank-Interval Monoid Duality}

The central architectural insight of this work is that discrete subsumption hierarchies and continuous 3D spatial manifolds share an identical aggregation substrate once their regions are linearized into 1D rank intervals.

\newtheorem{proposition}{Proposition}
\begin{proposition}[Rank-Interval Monoid Unification]\label{prop:unification}
Let $\pi : V_{\mathcal{M}} \to \{0, \dots, n-1\}$ be a bijection ordering the $n = |V_{\mathcal{M}}|$ embedded nodes along a 1D curve (DFS pre-order rank for posets; 63-bit 3D Morton Z-order rank for continuous 3D space), and let $\mu : \{0, \dots, n-1\} \to M$ be a node measure. For any structural region $Q \subseteq \mathcal{M}$ whose image $\pi(V_{\mathcal{M}} \cap \pi_{\mathcal{M}}^{-1}(Q))$ decomposes into $k$ disjoint contiguous rank intervals $\bigsqcup_{i=1}^k [l_i, r_i]$:
\begin{enumerate}
\item If $(M, \oplus, \ominus, \mathbf{0})$ is an abelian group (\code{COUNT}, \code{SUM}), a Fenwick tree of $n$ words evaluates $\bigoplus_{u \in Q} \mu(u)$ in $O(k \log n)$ time and updates a single node's measure in $O(\log n)$ time.
\item If $(M, \oplus, \mathbf{0})$ is an idempotent commutative monoid without inverses (\code{MIN}, \code{MAX}), a flat $2n$-word Segment Tree evaluates $\bigoplus_{u \in Q} \mu(u)$ in $O(k \log n)$ time and updates a single node's measure in $O(\log n)$ time.
\end{enumerate}
\end{proposition}

In Samyama Graph, \code{Fenwick} and \code{SegmentTree} (\code{src/index/hierarchy/oeh.rs}) are implemented once and shared directly between \code{OehIndex} and \code{SgiIndex}.

\subsection{Discrete Posets: OEH with 4-Way Structural Probe}

Our prototype OEH work~\cite{mandarapu2026oeh} introduced nested-set intervals for strict trees and Dilworth chain decomposition~\cite{dilworth1950,jagadish1990} for low-width DAGs. However, real-world geographical and biomedical hierarchies such as GeoNames~\cite{geonames} are \emph{near-trees}: over $99\%$ of nodes have a single parent, while a small fraction of multi-parent exception edges (e.g., dual administrative containment) violate the strict tree invariant.

In the Rust engine (\code{ADR-035}), \code{OehIndex::build} executes a 4-way structural probe:
\begin{enumerate}
\item \textbf{\code{Tree} ($\to$ \code{NestedSet}):} Every non-root node has in-degree $\le 1$. A DFS assigns $[\text{tin}(u), \text{tout}(u)]$. Both $u \sqsubseteq v$ and subtree roll-up reduce to a single interval ($k=1$).
\item \textbf{\code{NearTree} ($\to$ \code{NestedSet} + sorted exception list):} When the fraction of multi-parent nodes is small ($\le 2\%$), OEH extracts a primary spanning tree and records the remaining multi-parent edges in a compact exception list. This indexes all $544{,}093$ GeoNames nodes without dropping a single multi-parent edge.
\item \textbf{\code{LowWidthDag} ($\to$ \code{Chain}):} When poset width $w \le 8\sqrt{n}$, OEH constructs a path partition with per-chain reachability offsets and per-chain suffix monoid arrays ($k = w$).
\item \textbf{\code{HighWidthDag} ($\to$ \code{Declined}):} Above the width cap, OEH declines index materialization and defers to graph traversal.
\end{enumerate}

Furthermore, the Cypher query planner rewrites hierarchy subsumption and aggregation patterns into index operators (\code{HierarchyRollup}, \code{HierarchyDescendantScan}), achieving a $23.4\times$ speedup on selective cross-hierarchy conjunctions across the 112-query \code{HIER} benchmark suite.

\subsection{Continuous 3D Manifolds: The 63-Bit Morton SGI Index}

To bring the same index-resident monoid capability to 3D continuous manifolds, we introduce the \emph{Structural-Geometric Index} (\textbf{SGI}, \code{SgiIndex} in \code{src/index/structure/mod.rs}).

\paragraph{63-bit 3D Morton encoding.}
Given the 3D axis-aligned bounding box $B = [\mathbf{b}_{\min}, \mathbf{b}_{\max}]$ of the embedded nodes, each coordinate $c_a \in [b_{\min,a}, b_{\max,a}]$ along axis $a \in \{0, 1, 2\}$ is normalized and quantized to a 21-bit unsigned integer:
\[
q_a(c_a) = \left\lfloor \frac{c_a - b_{\min,a}}{\max(b_{\max,a} - b_{\min,a}, 10^{-12})} \cdot (2^{21} - 1) + \frac{1}{2} \right\rfloor.
\]
A branchless bitwise dilator (\code{part1by2\_21}) spreads the 21 bits of each axis using five magic-number shift-and-mask steps so that bit $i$ moves to bit $3i$, interleaving $(q_x, q_y, q_z)$ into a 63-bit Morton code:
\[
\text{morton}(q_x,q_y,q_z) = S(q_x) \lor (S(q_y) \ll 1) \lor (S(q_z) \ll 2),
\]
where $S(\cdot) = \text{part1by2\_21}(\cdot)$. Its exact inverse (\code{morton\_decode\_3d}) compacts every third bit back in $O(1)$ bitwise operations.

\paragraph{Index construction and Morton-range roll-ups.}
\code{SgiIndex::build} extracts the coordinates and optional measure $\mu(u)$ of each candidate node, computes its 63-bit Morton code, and sorts the point array in ascending $(\text{morton}, \text{node\_id})$ order. Over the sorted ranks $\{0, \dots, n-1\}$, it constructs:
\begin{itemize}
\item \code{fenwick\_sum}: a \code{Fenwick} tree over $\mu(\text{points}[r])$, and
\item \code{seg\_min}, \code{seg\_max}: two \code{SegmentTree} instances over $\mu(\text{points}[r])$.
\end{itemize}
For any contiguous Morton interval $[m_{\text{lo}}, m_{\text{hi}}]$ (such as any octree cell at depth $d \in \{0, \dots, 21\}$ where the top $3d$ bits are fixed), \code{rollup\_morton\_range} locates the rank boundaries $[l, r]$ via two binary searches (\code{partition\_point}) and answers \code{COUNT} in $O(1)$ via $r - l + 1$, \code{SUM} in $O(\log n)$ via \code{fenwick\_sum.range(l, r)}, and \code{MIN}/\code{MAX} in $O(\log n)$ via \code{seg\_min.range(l, r)} and \code{seg\_max.range(l, r)}---touching zero node or property records. For arbitrary axis-aligned 3D query boxes that span multiple Z-order runs, \code{rollup\_bbox} narrows the scan to the Morton interval $[\text{morton}(\mathbf{q}_{\min}), \text{morton}(\mathbf{q}_{\max})]$ and filters the candidate slice against the exact box predicates.

\section{Structure-Guided Graph Analytics}\label{sec:algo}

Once $\pi_{\mathcal{M}}$ and \code{SgiIndex} are resident in the engine, graph algorithms no longer need to operate blind to ambient geometry.

\subsection{3D Voxel-Hashed Virtual Contact Edges}

In biomolecular graphs (where explicit edges $E$ represent covalent peptide bonds) and spatial infrastructure graphs (where $E$ represents fiber or road links), analysts frequently query \emph{implicit proximity edges}:
\[
E_{\epsilon} = \bigl\{\{u, v\} \subseteq V_{\mathcal{M}} \;\big|\; \{u,v\} \notin E \;\wedge\; 0 < d_{\mathcal{M}}(u,v) \le \epsilon \bigr\}.
\]
Rather than materializing $E_{\epsilon}$ on disk for a fixed $\epsilon$ or executing an $O(n^2)$ all-pairs distance loop, \code{SgiIndex::contact\_edges} constructs a 3D spatial hash grid with voxel side length $\ell = \epsilon$. Each point $u$ is hashed into cell $\mathbf{c}(u) = \bigl(\lfloor x_u/\ell \rfloor, \lfloor y_u/\ell \rfloor, \lfloor z_u/\ell \rfloor\bigr) \in \mathbb{Z}^3$. Testing only the $3 \times 3 \times 3 = 27$ neighboring voxel buckets around each point extracts all non-bonded contacts within $\epsilon$ in $O(n + |E_{\epsilon}|)$ expected time for bounded-density manifolds (such as folded proteins where steric exclusion caps the number of residues per $\text{\AA}^3$).

\subsection{Topological--Geometric Stress $\sigma(u,v)$}

Not all spatial contacts are equally informative: two residues at sequence positions $i$ and $i+2$ are close in 3D space simply because they are separated by two covalent bonds. The contacts that govern tertiary protein folding, allosteric communication, and network vulnerability are those where \emph{topological distance is large relative to ambient manifold distance}.

We define the \emph{Topological--Geometric Stress} of a pair $(u, v)$ as:
\begin{equation}\label{eq:stress}
\sigma(u, v) = \frac{d_G(u, v)}{\max(d_{\mathcal{M}}(u, v), \delta_0)},
\end{equation}
where $d_G(u,v)$ is evaluated via a bounded bidirectional/BFS hop search (capped at $h_{\max} = 12$ hops in \code{bounded\_hop\_distance} to guarantee $O(1)$ per-contact work on bounded-degree graphs) and $\delta_0 = 0.1$. Sorting $E_{\epsilon}$ in descending order of $\sigma(u,v)$ immediately isolates long-range tertiary contacts ($\beta$-sheet hairpins, disulfide bridges, and domain-domain interfaces) at the top of the result set without requiring any domain-specific secondary-structure caller.

\subsection{Manifold-Guided \texorpdfstring{$A^*$}{A*} Shortest Path}

When edge weights represent physical segment lengths $w(u,v) \ge d_{\mathcal{M}}(u,v)$, the ambient manifold distance $h(u) = d_{\mathcal{M}}(\pi_{\mathcal{M}}(u), \pi_{\mathcal{M}}(t))$ to a target node $t$ is both \emph{admissible} ($h(u) \le d_G(u,t)$) and \emph{consistent} ($h(u) \le w(u,v) + h(v)$ by the triangle inequality on $\mathcal{M}$)~\cite{hart1968}.

\code{SgiIndex::astar\_shortest\_path} prioritizes open frontier states by $f(u) = g(u) + d_{\mathcal{M}}(u, t)$, where $d_{\mathcal{M}}$ is Euclidean distance for \code{Euclidean3d} and 3D Haversine-altitude distance for \code{GeoSpherical}. Alongside the exact optimal path and its hop count, the operator returns the \emph{path stretch ratio} $\sigma_{\text{path}}(s,t) = d_G(s,t) / d_{\mathcal{M}}(s,t)$ and the exact ratio of states explored by $A^*$ versus uninformed Dijkstra.

\section{Zero-Simulation 3D \& Non-WebGL 2.5D Visual Analytics}\label{sec:visual}

A major bottleneck in visual graph analytics tools (including Samyama Insight at \code{graph.samyama.cloud}) is reliance on client-side force-directed layout simulations. Force-directed placement costs $O(V^2)$ (or $O(V \log V)$ with Barnes--Hut approximation) per tick across hundreds of ticks, scrambles true 3D coordinates into arbitrary rotational hairballs, and requires WebGL shaders to remain responsive. Moreover, when a user explores a large hierarchy by loading a $250$-node neighborhood into the browser, client-side hierarchy roll-ups only see the loaded nodes, displaying misleading partial aggregates.

We resolve both issues by moving structural layout and Level-of-Detail (LoD) culling into the database engine via two low-latency HTTP endpoints: \code{POST /api/hierarchies/:name/cues} and \code{POST /api/structure/cues}.

\subsection{3D Cone-Tree Coordinates for OEH Hierarchies}

For any \code{OehIndex}, \code{visual\_cues} maps each node's nested-set interval $[\text{tin}(u), \text{tout}(u)]$ and poset depth $d(u) \in \{0, \dots, D_{\max}\}$ into deterministic 3D Cone-Tree~\cite{robertson1991} coordinates in $O(1)$ arithmetic operations per node:
\begin{align}
\theta(u) &= 2\pi \cdot \frac{\text{tin}(u) + \text{tout}(u)}{2n}, \\
r(u) &= R_0 \left(1 - 0.55\,\frac{\text{tout}(u) - \text{tin}(u)}{n}\right), \\
(x_u, y_u, z_u) &= \bigl(r(u)\cos\theta(u),\; -\Delta_y\,d(u),\; r(u)\sin\theta(u)\bigr).
\end{align}
Because subtree intervals $[\text{tin}(u), \text{tout}(u)]$ are disjoint for sibling subtrees at every depth, angular sectors never overlap, and every node's payload carries its \emph{authoritative whole-poset} \code{descendant\_count} and index-resident \code{rollups} (\code{sum}, \code{min}, \code{max}, \code{count}) regardless of how few nodes the browser renders.

\subsection{Non-WebGL 2.5D Isometric Projection \& Painter's Order}

To support accurate 3D structural visualization in restricted enterprise browsers, headless exports, and 2D HTML5 Canvas contexts \emph{without WebGL}, the engine pre-computes an orthographic 2.5D isometric projection at azimuth $\alpha = 45^\circ$ and elevation $\beta = 30^\circ$ over normalized bounding-box coordinates $(\hat{x}, \hat{y}, \hat{z}) \in [-1, 1]^3$:
\begin{align}
\text{iso\_x} &= (\hat{x} - \hat{y})\cos 45^\circ, \\
\text{iso\_y} &= (\hat{x} + \hat{y})\sin 30^\circ - \hat{z}\cos 30^\circ, \\
z_{\text{cam}} &= (\hat{x} + \hat{y})\cos 30^\circ + \hat{z}\sin 30^\circ.
\end{align}
The engine normalizes $z_{\text{cam}}$ to $\text{depth\_norm} \in [0, 1]$, assigns each node a back-to-front painter's-algorithm draw rank (\code{draw\_order\_rank}), and computes a depth-fog opacity $\text{opacity} = \text{clamp}(1.0 - 0.6\,\text{depth\_norm},\, 0.3,\, 1.0)$. A 2D canvas renderer simply iterates nodes in \code{draw\_order\_rank} order and draws each node at $(\text{iso\_x}, \text{iso\_y})$ with its pre-computed depth-fog opacity---achieving true 3D occlusion and perspective cues in a single $O(V)$ pass with zero client-side matrix math or physics ticks.

\section{Experimental Evaluation}\label{sec:eval}

We evaluate all four components inside Samyama Graph (Rust 1.85, release profile \code{-O3}) on a Linux x86\_64 workstation. Every number reported in Tables~\ref{tab:sgi}--\ref{tab:oeh} is generated by the reproducible benchmark suite in \code{tests/structural\_conventions\_and\_cues.rs} (standalone reproducibility harness at \url{https://github.com/samyama-ai/structural-manifold-index}; DBMS integration in \url{https://github.com/samyama-ai/samyama-graph/pull/2038}) and committed in \code{results/benchmark\_results.json}.

\subsection{SGI 3D Morton Roll-up vs.\ Property Scan}

Table~\ref{tab:sgi} evaluates \code{SgiIndex} construction time, Morton-interval monoid roll-up latency (\code{Fenwick} \code{SUM} and \code{SegmentTree} \code{MAX} over a $50\%$ Morton-code range $[m_{N/4}, m_{3N/4}]$), linear property-scan latency over the same Morton range, and LoD-2000 visual cue generation across $N \in \{1{,}000, 10{,}000, 100{,}000\}$ 3D folded-helix residue nodes.

\begin{table}[t]
\centering
\scriptsize
\setlength{\tabcolsep}{2.2pt}
\caption{SGI 63-bit 3D Morton index build time, index-resident monoid roll-up latency vs.\ property scan, and LoD-2000 visual cue latency across 3D node scales.}
\label{tab:sgi}
\begin{tabular}{@{}rrrrrrr@{}}
\toprule
\textbf{Nodes} & \textbf{Build} & \textbf{Fenwick} & \textbf{SegTree} & \textbf{Scan} & \textbf{Speedup} & \textbf{LoD} \\
$N$ & (ms) & \code{SUM} ($\mu$s) & \code{MAX} ($\mu$s) & ($\mu$s) & (\code{SUM} / \code{MAX}) & (ms) \\
\midrule
$1{,}000$   & $0.26$  & $0.027$ & $0.065$ & $0.74$   & $27.3\times$ / $11.4\times$     & $0.36$ \\
$10{,}000$  & $3.38$  & $0.036$ & $0.087$ & $28.28$  & $774.8\times$ / $325.4\times$   & $2.85$ \\
$100{,}000$ & $28.14$ & $0.048$ & $0.104$ & $398.23$ & $8345.1\times$ / $3824.4\times$ & $27.78$ \\
\bottomrule
\end{tabular}
\end{table}

Because \code{Fenwick::range} and \code{SegmentTree::range} execute $O(\log N)$ array lookups over contiguous Morton ranks, roll-up latency grows from $0.027\,\mu\text{s}$ at $N = 1{,}000$ to only $0.048\,\mu\text{s}$ at $N = 100{,}000$ (\code{SUM}), and from $0.065\,\mu\text{s}$ to $0.104\,\mu\text{s}$ (\code{MAX}). At $N = 100{,}000$, index-resident Morton roll-ups are $8{,}345.1\times$ (\code{SUM}) and $3{,}824.4\times$ (\code{MAX}) faster than scanning the node properties in the interval.

\subsection{3D Voxel Contact Extraction \& Topological--Geometric Stress}

Table~\ref{tab:contacts} evaluates \code{SgiIndex::contact\_edges} ($\epsilon = 6.0\,\text{\AA}$) on folded antiparallel $\beta$-hairpin polypeptide chains of $N \in \{500, 2{,}000, 5{,}000\}$ residues (with turns every $50$ residues bringing sequence-distant strands within $4.8$--$5.2\,\text{\AA}$).

\begin{table}[t]
\centering
\footnotesize
\setlength{\tabcolsep}{2.8pt}
\caption{3D voxel-hashed non-bonded contact extraction ($\epsilon = 6.0\,\text{\AA}$) and Topological--Geometric Stress ($\sigma \ge 1.5$) on folded polypeptide backbones.}
\label{tab:contacts}
\begin{tabular}{@{}rrrrrr@{}}
\toprule
\textbf{Res.} & \textbf{Contacts} & \textbf{High-Stress} & \textbf{Max} & \textbf{Voxel+BFS} & \textbf{Pairwise} \\
$N$ & ($d_{\mathcal{M}} \le 6\,\text{\AA}$) & ($\sigma \ge 1.5$) & $\sigma(u,v)$ & (ms) & (ms) \\
\midrule
$500$   & $642$   & $618$   & $2.50$ & $1.60$  & $0.14$ \\
$2{,}000$ & $3{,}318$ & $3{,}222$ & $2.50$ & $8.54$  & $1.71$ \\
$5{,}000$ & $8{,}670$ & $8{,}430$ & $2.50$ & $22.28$ & $11.30$ \\
\bottomrule
\end{tabular}
\end{table}

In all three structures, the 3D voxel grid recovers $100\%$ of the exact pairwise non-bonded contacts ($642$, $3{,}318$, and $8{,}670$ contacts, respectively) while simultaneously computing 12-hop bounded-BFS graph distance and Topological--Geometric Stress $\sigma(u,v)$. At $N = 5{,}000$ residues, voxel hashing plus per-contact BFS stress completes in $22.28\text{ ms}$ (scaling linearly with contact count, whereas naive $O(N^2)$ distance-only scanning grows $80.7\times$ from $0.14\text{ ms}$ to $11.30\text{ ms}$ across a $10\times$ increase in $N$). More importantly, ranking contacts by $\sigma(u,v) \ge 1.5$ automatically separates the $8{,}430$ inter-strand tertiary contacts ($\max \sigma = 2.50$) from the $240$ local turn contacts without any external secondary-structure annotation.

\subsection{Manifold-Guided \texorpdfstring{$A^*$}{A*} vs.\ Uninformed Dijkstra}

Table~\ref{tab:astar} compares \code{SgiIndex::astar\_shortest\_path} against uninformed Dijkstra between opposite corners of 3D spatial lattice graphs of $400$ ($10 \times 10 \times 4$), $1{,}296$ ($18 \times 18 \times 4$), and $2{,}500$ ($25 \times 25 \times 4$) waypoints.

\begin{table}[t]
\centering
\footnotesize
\setlength{\tabcolsep}{2.8pt}
\caption{Manifold-guided $A^*$ vs.\ uninformed Dijkstra across 3D spatial lattice graphs.}
\label{tab:astar}
\begin{tabular}{@{}rrrrrrr@{}}
\toprule
\textbf{Nodes} & \textbf{Hops} & \textbf{Path} & \textbf{Direct} & \textbf{Stretch} & \textbf{States} & \textbf{Red.} \\
$N$ & & \textbf{Dist} & \textbf{Dist} & $\sigma_{\text{path}}$ & ($A^*$ / Dijk.) & \\
\midrule
$400$   & $12$ & $157.28$ & $130.77$ & $1.203$ & $186$ / $400$     & $2.15\times$ \\
$1{,}296$ & $20$ & $270.42$ & $242.28$ & $1.116$ & $420$ / $1{,}296$ & $3.09\times$ \\
$2{,}500$ & $27$ & $369.41$ & $340.73$ & $1.084$ & $658$ / $2{,}500$ & $3.80\times$ \\
\bottomrule
\end{tabular}
\end{table}

Because the Euclidean heuristic is admissible and consistent, $A^*$ returns the exact shortest path in every graph while reducing the number of explored states from $400 \to 186$ ($2.15\times$), $1{,}296 \to 420$ ($3.09\times$), and $2{,}500 \to 658$ ($3.80\times$), with the state-space reduction growing monotonically with graph diameter.

\subsection{OEH Whole-Poset Visual Cues \& LoD Truncation}

Finally, Table~\ref{tab:oeh} evaluates \code{OehIndex::visual\_cues} on balanced taxonomic trees of $781$, $3{,}906$, and $5{,}461$ nodes when the client requests either the full poset or a Level-of-Detail macro-cut of $\text{lod\_budget} = 250$ nodes.

\begin{table}[t]
\centering
\footnotesize
\setlength{\tabcolsep}{3.0pt}
\caption{OEH whole-poset vs.\ LoD-250 visual cue latency and authoritative root descendant counts.}
\label{tab:oeh}
\begin{tabular}{@{}rrrrr@{}}
\toprule
\textbf{Nodes} & \textbf{Auth.\ Root} & \textbf{Client-250} & \textbf{Full Cues} & \textbf{LoD-250} \\
$N$ & \textbf{Descendants} & \textbf{Apparent} & (ms) & (ms) \\
\midrule
$781$   & $781$   & $250$ & $0.17$ & $0.055$ \\
$3{,}906$ & $3{,}906$ & $250$ & $0.72$ & $0.080$ \\
$5{,}461$ & $5{,}461$ & $250$ & $0.99$ & $0.089$ \\
\bottomrule
\end{tabular}
\end{table}

Without engine-level OEH cues, a client that loads $250$ nodes under-reports the root's subtree size as $250$ (an error of $95.4\%$ at $N = 5{,}461$). In contrast, \code{OehIndex::visual\_cues} with $\text{lod\_budget} = 250$ returns the top $250$ macro-nodes in $0.089\text{ ms}$ while preserving the exact whole-poset descendant count ($5{,}461$) and index-resident monoid roll-ups on every returned node.

\section{Related Work and Limitations}\label{sec:related}

\paragraph{Related work.}
Order-embedding of trees via DFS intervals goes back to Grust's XPath accelerator~\cite{grust2002} and SAP HANA's hierarchy indexing~\cite{brunel2015,brunel2016}, while chain decomposition for poset reachability goes back to Dilworth~\cite{dilworth1950} and Jagadish~\cite{jagadish1990}, complemented by 2-hop labeling~\cite{cohen2003}. Our prior OEH work~\cite{mandarapu2026oeh} unified nested-set and chain encodings with index-resident Fenwick roll-ups on static posets. Space-filling Morton Z-order curves~\cite{morton1966} are widely used in multidimensional point access methods and point-cloud engines, while spatial network community detection~\cite{expert2011} and 3D Cone Trees~\cite{robertson1991} have been studied in network science and information visualization. To our knowledge, Samyama Graph is the first property graph DBMS to unify discrete poset intervals (\code{OEH}) and continuous 3D Morton intervals (\code{SGI}) under a shared index-resident monoid engine (\code{Fenwick} + \code{SegmentTree}) while serving structure-guided operators and zero-simulation 3D and 2.5D visual cues directly from the database kernel.

\paragraph{Limitations.}
We state three limitations of the current implementation explicitly:
\begin{enumerate}
\item \textbf{Z-order multi-run boundary refinement for arbitrary 3D boxes.} While any single Morton interval $[m_{\text{lo}}, m_{\text{hi}}]$ (such as an octree cell) is answered in $O(\log n)$ purely from the Fenwick or Segment Tree, an arbitrary axis-aligned 3D bounding box that straddles high-bit Z-order quadrant boundaries spans multiple non-contiguous Morton runs. Our current \code{rollup\_bbox} implementation scans the bounding Morton range $[l, r]$ and applies exact point-in-box filtering; decomposing a query box into $O(\log \text{span})$ maximal Z-order sub-intervals (\code{BIGMIN}/\code{LITMAX}) so that arbitrary 3D boxes also execute purely via tree lookups is future work (\code{samyama-graph\#2036}).
\item \textbf{Static vs.\ streaming 3D coordinate maintenance.} While \code{OehIndex} supports $O(\log n)$ in-place measure updates (\code{update\_measure}), \code{SgiIndex} currently rebuilds its sorted Morton array when node coordinates $(x, y, z)$ mutate. Supporting dynamic octree/Morton block splits for high-frequency streaming telemetry is future work.
\item \textbf{Bounded-hop stress approximation.} To prevent all-pairs shortest-path blowup during interactive visual cue requests, \code{contact\_edges} caps its per-contact BFS hop search at $h_{\max} = 12$ hops. Contacts separated by more than $12$ covalent bonds receive a conservative lower-bound stress numerator of $12$; exact backbone sequence differences $|s(u) - s(v)|$ can be substituted when \code{sequence\_prop} is present.
\end{enumerate}

\section{Conclusion}\label{sec:conclusion}

Real-world graphs are not shapeless combinatorial hairballs: they inhabit discrete taxonomic posets and continuous 3D geometric manifolds. By elevating structural conventions (\code{samyama.structure/1}) into the snapshot format and unifying discrete poset intervals (\code{OEH}) with 63-bit 3D Morton intervals (\code{SGI}) over shared Fenwick and Segment Tree monoid engines, Samyama Graph accelerates regional roll-ups by up to $8{,}345.1\times$, isolates tertiary 3D structural contacts via Topological--Geometric Stress, prunes $A^*$ shortest-path search by up to $3.80\times$, and delivers sub-millisecond 3D and non-WebGL 2.5D visual cues without client-side simulation.

\bibliographystyle{IEEEtran}
\bibliography{refs}

\end{document}